\documentclass[conference,10pt]{IEEEtran}

\usepackage{hyperref}	
\usepackage{setspace}
\usepackage{times}
\usepackage{helvet}
\usepackage{courier}
\usepackage{enumitem}
\usepackage{graphicx}
\usepackage{graphics}
\usepackage{boiboites}
\usepackage{color}
\usepackage{subfigure}

\usepackage{floatrow}
\usepackage{multirow}
\usepackage{xcolor}
\usepackage{grffile}
\usepackage{tabularx,colortbl}
\usepackage{float} 
\usepackage{balance}
\usepackage{moresize}
\usepackage{amssymb,amsmath,amsthm}
\usepackage{mathtools} 

\usepackage{flushend}
\usepackage{xspace}

\usepackage{amssymb}
\usepackage{pifont}
\newcommand{\cmark}{\ding{51}}%
\newcommand{\xmark}{\ding{55}}%

\theoremstyle{definition}

\newtheorem{problem}{Problem}

\newtheorem*{infproblem*}{Informal~Problem}

\newtheorem{definition}{Definition}

\newcommand{\hide}[1]{}
\newcommand{\method}{MIMiS\xspace}

\newcommand{\tensor}[1]{\underline{\mathbf{#1} }}

\usepackage[ruled,linesnumbered]{algorithm2e}
\usepackage{algorithmic}

\newcounter{ALC@tempcntr}

\makeatletter
\def\@IEEEpubidpullup{8\baselineskip}
\makeatother

\begin{document}

\title{\method: Minimally Intrusive Mining of Smartphone User Behaviors}
\author{
Pravallika Devineni$^1$, Evangelos E. Papalexakis$^1$, Kalina Michalska$^2$ and Michalis Faloutsos$^1$ \\ $^1$Department of Computer Science, University of California, Riverside, CA \\ $^2$Department of Psychology, University of California, Riverside, CA \\
pdevi002@ucr.edu \hspace{1.5em}
\{epapalex, michalis\}@cs.ucr.edu \hspace{1.5em}
kalinam@ucr.edu
}

\maketitle

\begin{abstract}

How intrusive does a life-saving user-monitoring application really need to be? While most previous research was focused on analyzing mental state of users from social media and smartphones, there is little effort towards protecting user privacy in these analyses. A challenge in analyzing user behaviors is that not only is the data multi-dimensional with a myriad of user activities but these activities occur at varying temporal rates. The overarching question of our work is: Given a set of sensitive user features, what is the minimum amount of information required to group users with similar behavior? Furthermore, does this user behavior correlate with their mental state? 
Towards answering those questions, our contributions are two fold: we introduce the concept of privacy surfaces that combine sensitive user data at different levels of intrusiveness.  As our second contribution, we introduce \method, an unsupervised privacy-aware  framework that clusters users in a given privacy surface configuration to homogeneous groups with respect to their temporal signature. In addition, we explore the trade-off between intrusiveness and prediction accuracy. \method employs multi-set decomposition in order to deal with incompatible temporal granularities in user activities.
We extensively evaluate \method  on real data. Across a variety of privacy surfaces, \method identified groups that are highly homogeneous with respect to self-reported mental health scores. Finally, we conduct an in-depth exploration of the discovered clusters, identifying groups whose behavior is consistent with academic deadlines.

\end{abstract}

\section{Introduction}
\label{sec:intro}

Online social networks and smartphones produce a vast amount of user-generated content providing opportunities to study human behavior and social interactions at scale. On the one hand, we have an abundance of online activity data, whose careful mining could improve the lives of users. On the other hand, there are grave concerns about the intrusiveness of such data collection applications. For example, a student struggling with depression could greatly benefit from an early intervention by their parents or advisors. At the same time, where do we draw the line of what we are allowed to monitor? A key question in this complex socio-technical conundrum is to find the minimum amount of information needed for accurate and timely detection.

Most studies in the area of human behavior data mining focus on online behavior and mental health \cite{DeChoudhury14PPD, Coppersmith2014quantifying}, personality mining from online behavior \cite{Kosinski13traits, Murnane2014unraveling}, and analysis of large user networks \cite{Viswanath09FB}. In this paper, we focus on capturing user behavioral patterns using information that is minimally invasive to privacy from smartphone sensors to predict well-being, mental state, and destructive behaviors. In wake of the Cambridge Analytica scandal \cite{CA2019scandal}, where several millions of Facebook user data was exchanged between companies compromising user privacy, we believe this is an important question to explore and understand. 

\begin{figure}[t!]
	\begin{center}
		\includegraphics[width=1\textwidth]{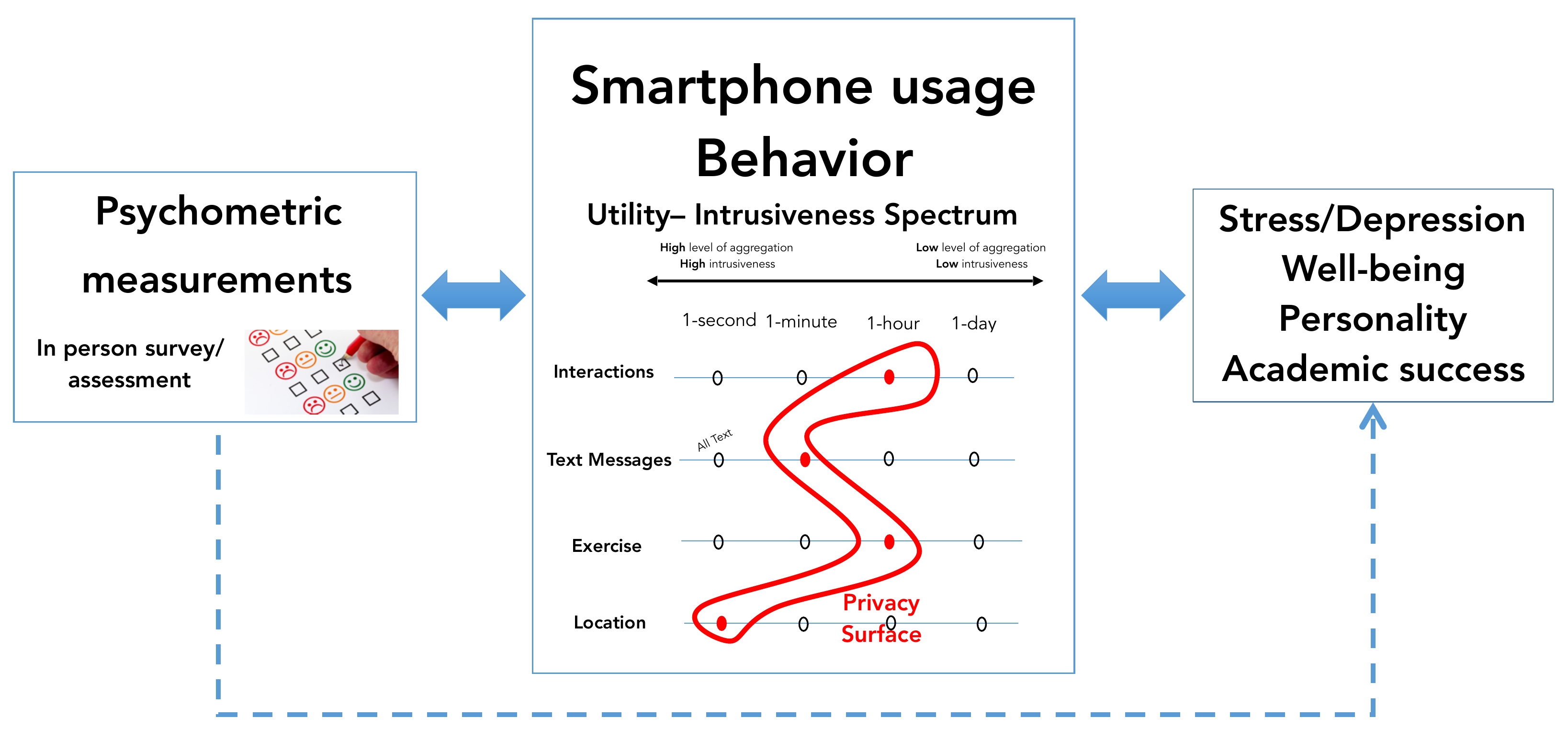}
		\vspace{-0.32in}
		\caption{Our overarching goal is to identify the minimum amount of privacy-invasive information from online user activities in order to predict well-being and mental state. We propose the idea of {\bf privacy surfaces} that combines user activity data at different temporal granularities and explore the trade-off between protecting user privacy and achieving prediction accuracy.}
		\label{fig:crown}
	\end{center}
\end{figure}

Off-the-shelf smartphones come equipped with various sensors, thus providing us the capability to continuously collect user data. Sensors like accelerometer, geolocation and WiFi help us infer a person's physical movements while microphone activity, phone calls and text messages might provide an insight into a person's mental state. The fundamental questions that we are asking in this work is {\em can we use such smartphone behaviors as proxies for estimating a user's mental state and well-being?}; and if so, {\em what is the minimum amount of sensitive or intrusive information that we need, in order to achieve that?}

As exciting as it is to study human behavior, sensor information in its raw form is privacy-intrusive and compromises the safety of the user. To tackle the issue of privacy with regards to user data, we introduce the concept of {\em privacy surfaces}. Privacy surfaces combine user sensor information at varying levels of intrusiveness, where we define intrusiveness based on the temporal granularity of the information. Figure \ref{fig:crown} demonstrates an exemplary privacy surface and how it fits into answering the fundamental questions we pose.

Furthermore, towards achieving minimally invasive smartphone user behavior mining, and discovering how it correlates with mental well-being, we propose \method,  a privacy-aware unsupervised framework which is based on multi-set decomposition, an advanced generalization of tensor decomposition. \method takes as input a) smartphone data for a set of users and b) a privacy surface configuration, defining the levels of intrusiveness of the data, and produces groups of users that exhibit similar behavior over time. Furthermore, after extensive experimentation, we demonstrate that the clustering of users by \method correlates very well with self-reported mental states, such as depression, loneliness, and stress.

We summarize our contributions as follows:
\begin{itemize}
    \item {\bf Novel framework}: We propose \method, a privacy-aware framework that extracts patterns from smartphone sensor data in an unsupervised fashion. 
    \item {\bf Exploratory Analysis}: We extensively evaluate the underlying properties of the discovered clusters. \method groups users into clusters which are highly homogeneous with respect to different measures of mental health. For example, those groups with high academic performance, low physical activity and high levels of stress. 
    \item {\bf Validation against ground truth}: We associate the discovered latent temporal patterns against the ground truth (in the form of psychometric scale measures) to demonstrate the usefulness of our approach.
\end{itemize}

\section{Dataset and Problem Definition}
\label{sec:problem}

\method is a general framework that can be applied to a variety of smartphone user behavior datasets. In order to ground \method to a real case study, in the remainder of the paper we will be using the StudentLife dataset \cite{Wang2014studentlife} as our running example.
This section presents the dataset and some useful definitions.

\subsection{Dataset}

We use the StudentLife dataset \cite{Wang2014studentlife}, a large publicly available dataset, tracking student performance, well-being and psysiological state. StudentLife is a 10-week study conducted on 48 Dartmouth students during 2013 spring quarter, among which are 30 undergraduates and 18 graduate students, and 10 female and 38 male students, in terms of gender. The dataset consists of four parts: smartphone sensors, ecological momentary assessments (EMAs), psychometric surveys, and academic performances. The smartphone sensor data include GPS, WiFi, accelerometer, call log, SMS, conversation, audio, phone lock, phone charge, and darkness. \hide{They are provided as metadata the feature instances and their timestamps.} The physical activity (stationary, walking, running, and unknown) and audio (silence, voice, noise, and unknown) were inferred from accelerometer and audio sensors respectively. This continual collection of sensor data provides us with capabilities to infer daily activities like mobility, sociability, sleep, and exercise. 

Psychometric scales measure knowledge, abilities, attitudes, and personality traits and those we use in this paper are presented in Table~\ref{tab:surveys}. \hide{The psychometrics collected pre and post survey include Big Five personality \cite{John1999bigfive}, flourishing scale \cite{Diener2010fs}, UCLA loneliness scale \cite{Russell1978ls}, positive and negative affect schedule \cite{Watson1988panas}, perceived stress scale (PSS) \cite{Cohen1983pss}, PHQ-9 depression scale \cite{Kroenke2002phq}, Pittsburgh sleep quality index (PSQI) \cite{Buysse1989sleep} and VR-12 health scale \cite{Kazis2006vr12}.} All 48 students answered the pre-survey questionnaires, while only 41 answered the post-survey. We replace the missing post-survey data with the pre-survey measures and drop those students who haven't answered either of the surveys from our study. The EMAs are scheduled daily questions to collect user mood in-situ. However, we do not use them in this work. The academic performance information includes class schedule, overall GPA, deadlines per day, and Piazza online forum participation for class. 

\begin{table}[t]
    \begin{tabular}{|p{3.5cm}|p{3.2cm}|}
    \hline
    {\bf Psychometric survey} & {\bf Measure} \\ 
    \hline
    Patient Health Questionnaire (PHQ-9) \cite{Kroenke2002phq} & depression level \\
    \hline
    Perceived Stress Scale (PSS) \cite{Cohen1983pss} & stress level \\
    \hline 
    Flourishing Scale \cite{Diener2010fs} & flourishing level \\
    \hline
    UCLA Loneliness scale \cite{Russell1978ls} & loneliness level \\
    \hline
    PANAS \cite{Watson1988panas} & positive and negative emotions \\
    \hline
    \end{tabular}
    \caption{Mental well-being surveys.}
    \label{tab:surveys}
\end{table}

\subsection{Research Problem}

The StudentLife dataset provides deep insight into the everyday life of smartphone users. Such sensitive data needs to be dealt with care not compromising the safety and privacy of users, while providing accurate predictions. Some features are more privacy-intrusive than others. For example, user geolocation consists of the latitude and longitude pinpointing the exact location at a given timestamp. We define the research question as follows: 

\begin{problem}\label{prob:formal_problem}
{\bf Given} a set of sensitive user features, {\bf find}  the minimum amount of information that can enable us to identify meaningful user clusters.
\end{problem}

To address the above problem, we need to determine two things 1) what information is more sensitive than others? and 2) what information should we use to achieve high-quality clustering? In the context of privacy awareness for sensitive user data, we define the follow terms:

\begin{definition}\label{def:modality} {\bf Modality of Information:} Modality distinguishes the aspects of privacy such as content (what the user is saying), network of interactions (who the user interacts with), and geolocation (physical location at a given time).
\end{definition}

\begin{definition}\label{def:privacy level} {\bf Privacy level:} The granularity or privacy level of a modality defines the level of temporal detail we have about the user data with respect to that modality.
\end{definition}

In the {\em StudentLife } dataset, modality refers to sensor data like accelerometer, GPS etc and privacy level refers to the time granularity of the data. We use the words sensor data and features interchangeably in this paper. For example, geolocation has higher modality than duration of phone being locked. Smartphones allow us to collect data at a very fine temporal granularity e.g. per-second instances. Data with coarse temporal granularity has higher privacy level and is less intrusive than fine per-second timestamped data.

\begin{definition}\label{def:privacy surface} {\bf Privacy surface:} A privacy surface is a combination of modalities at different privacy levels.
\end{definition}

\begin{definition}\label{def:utility} {\bf Utility: }
Utility refers to whether the information at that privacy level is sufficient to enable an effective algorithm to group users with similar behaviors.
\end{definition}

\begin{definition}\label{def:utlity_intrusiveness} {\bf Utility-Intrusiveness Spectrum: } The utility-intrusiveness spectrum refers to the interplay between the privacy level of a modality and the associated utility.
\end{definition}

To address Problem~\ref{prob:formal_problem}, we need to create privacy surfaces to discover user clusters, while exploring the trade-offs between the utility-intrusiveness space and the prediction accuracy. The homogeneity of clusters is determined with respect to the mental health states of users in each cluster. \hide{In addition, the discovered clusters relate to different mental health states of users. We determine homogeneity of the clusters with respect to the mental health states of users in each cluster.}

\subsection{Feature Extraction}

We extracted 18 features from 11 smartphone sensors in the StudentLife dataset, spanning across 66 days. We aggregated the features at five temporal granularities - 1-minute, 15-minute, 30-minute, 1-hour, and 1-day time bins. \hide{Each temporal granularity is time-bin. For example, 66 days has 1584 1-hour time bins.} The physical activity is inferred at 1-minute aggregation - stationary, walking, running, unknown, and the number of changes in activity per time bin. An activity change occurred if a user moves from stationary to walking or similar. For example, in a 1-hour time bin, a user was stationary for 48 minutes and was running for 12 minutes. The audio inferences are silence, voice, noise, and unknown. For bluetooth and WiFi, we count the number of unique encounters. For GPS, we count the number of unique locations (latitude and longitude) user visited in that time bin. We count the number of calls and SMS' recorded in each bin. We compute the number of minutes per time-bin for the features, conversation, darkness, phone charge, and phone lock. The extracted features could be one of the two types - minutes per bin and count of occurrences per bin. To normalize the values in the former, we divide the minutes by the number of minutes in the time bin. For features with counts, we logarithmically scale the data since log transformation spreads them over a wider range, while keeping the higher values compressed.

\section{Background}
\label{sec:method}

In numerous research areas such social networks \cite{Acar2005chatroom}, neuroscience \cite{Estienne2001multi}, and text mining \cite{Chew2007lang}, that underlying information content of the data may not be captured accurately by two-way data analysis. Such data can be represented as multi-way arrays, most popularly called {\em tensors}. Tensors are multi-dimensional extensions of matrices. Because of their ability to express multi-modal data, they are very powerful tools in applications that inherently create such data.

\begin{table}[t]
\label{tab:dfn}
\caption{Symbols and Definitions.}
\begin{tabular}{|c|l|}
\hline
\textbf{Symbol}     & \textbf{Definition }\\ 
\hline 
$\tensor{X},\mathbf{X}$ & Tensor or Multi-set, Matrix  \\
$\mathbf{x},x$ & Column vector, Scalar \\
$\mathbb{R}$ & Set of Real Numbers  \\ 
$\circ$ & Outer product  \\ 
\hide{$\mathbf{x}(I)$ & Spanning the elements of $\mathbf{x}$ in indices $\in I$ \\ 
$\mathbf{x}(:)$ & Spanning all elements of $\mathbf{x}$\\ 
$\mathbf{X}(:,r)$ &$ r^{th}$ column of $\mathbf{X}$  \\ 
$\mathbf{X}(r,:)$ & $ r^{th}$ row of $\mathbf{X}$  \\ }
\hline 
\end{tabular} 
\end{table}

\subsection{Tensors: Preliminaries}

\begin{figure}
    \centering
    \includegraphics[width=1\textwidth]{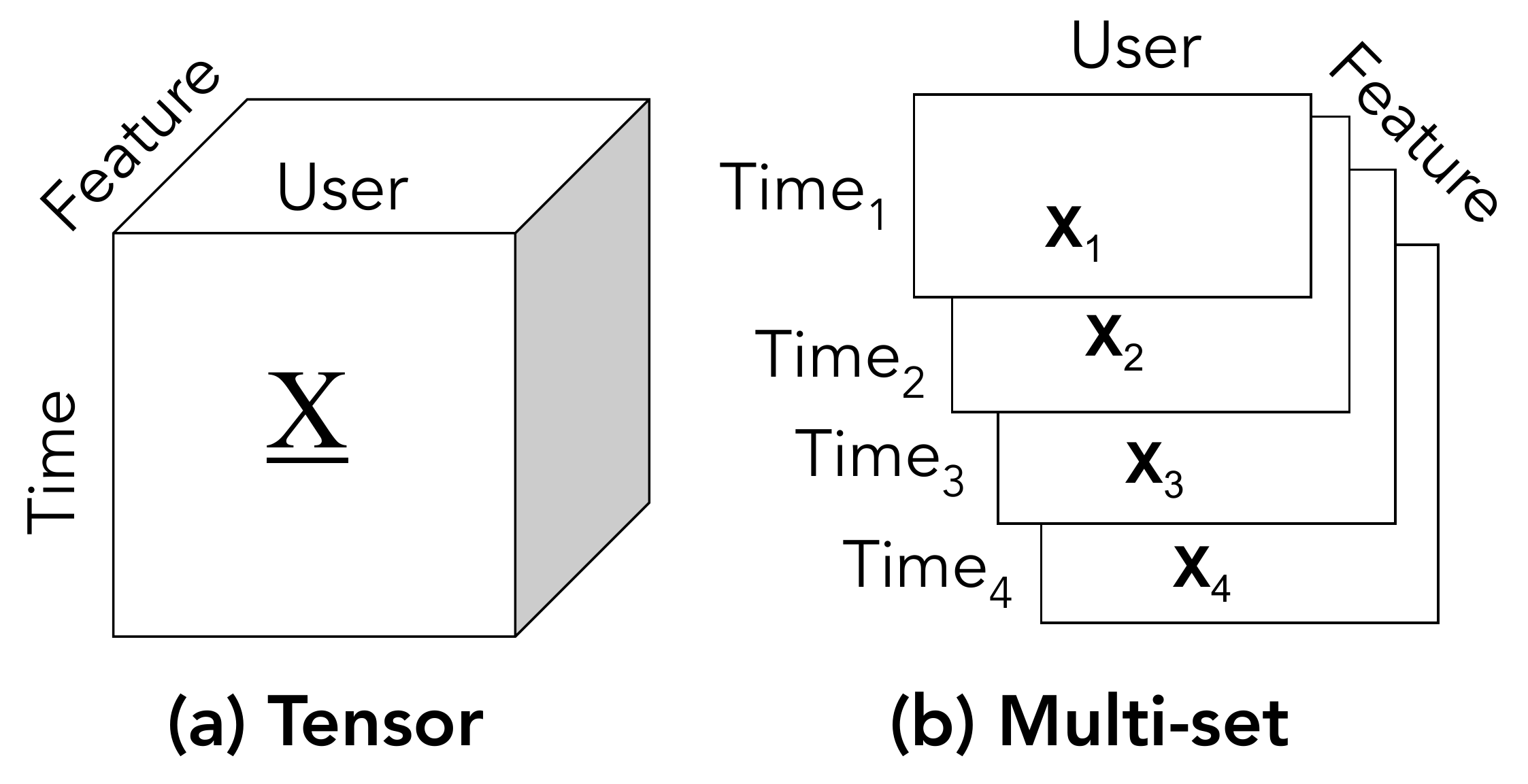}
    \caption{A tensor is collection of matrices that is trilinear across all modes. A multi-set is relaxed one mode. The varying mode in the multi-set  Time-User-Feature represented in (b) is Time.}
    \label{fig:tensor_vs_multiset}
\end{figure}

A tensor is an N-way array of data or objects. The {\em order} of a tensor denotes the number of its dimensions, also known as {\em ways} or {\em modes}. A scalar is a zeroth-order tensor; vector is a first-order tensor; and a matrix is a second-order tensor. Formally, an Nth-order tensor can be defined as: $\tensor{X} \in \mathbb{R}^{I_1 \times I_2 \times ... \times I_N}$, where the size of the $i$th dimension of $\tensor{X}$ is represented by $I_i$. A three-mode tensor is shown in Figure~\ref{fig:tensor_vs_multiset}(a). The two-dimensional sections of a higher-order tensor are called {\em slices}, while the higher-order counterparts of matrix columns and rows are called {\em fibers} of a tensor. When $\tensor{X}$ is sliced along the first dimension, horizontal slices of $\tensor{X}$ are formed. Similarly, slicing $\tensor{X}$ along second and third dimensions forms lateral and frontal slices respectively.\hide{, as shown in Figure~\ref{fig:tensor_slices}.} {\em Matricization} unfolds a tensor into a matrix in $N$ ways, one for each mode and is denoted by $\mathbf{X}_{(n)} \in \mathbb{R}^{I_n \times {I_1}{I_2}...{I_{n-1}}{I_{n+1}}...{I_N}}$.

\noindent {\em Tensor vs. multi-set:} A tensor does not allow for non-alignment among modes. Such data can be represented using a {\em multi-set}. A multi-set is a collection of matrices $\{\mathbf{X}_k\}$ for $k = 1...K$ that have one mode in common. These matrices can be seen as nearly forming a tensor; however the non-shared mode has different dimensions. A representation of multi-set is presented in Figure~\ref{fig:tensor_vs_multiset}(b). \hide{For example, users across a social network interact differently and possess different temporal granularities. This could be represented as a multi-set (user-feature-time), where time mode varies per user and per feature.}

\subsection{Tensor Decomposition}

\noindent {\bf Canonical Polyadic}

One the most popular tensor decompositions is Canonical Polyadic or CP. In principle, CP is simply an extension of Principal Component Analysis (PCA) to higher order data. Mathematically, CP can be represented as the decomposition of rank-one tensors. An $N$th-order rank-one tensor is a tensor that can be written as the outer product of $N$ vectors. The CP decomposition of a third-order tensor $\tensor{X} \in \mathbb{R}^{I\times J \times K}$ is represented as follows: 

\begin{equation}  \label{eq:cp_deco}
  \tensor{X} \approx \sum\limits_{r=1}^{R} \mathbf{u}_r \circ \mathbf{v}_r \circ \mathbf{w}_r
\end{equation}

where $\mathbf{u}_r \in \mathbb{R}^I$, $\mathbf{v}_r \in \mathbb{R}^J$ and $\mathbf{w}_r \in \mathbb{R}^K$ are column vectors. If we assemble the column vectors $\mathbf{u}_r$, $\mathbf{v}_r$, $\mathbf{w}_r$ as: $\mathbf{U} = [\mathbf{u_1} \mathbf{u_2}...\mathbf{u_R}] \in \mathbb{R^{I \times R}}$, $\mathbf{V} = [\mathbf{v_1} \mathbf{v_2}...\mathbf{v_R}] \in \mathbb{R^{J \times R}}$, $\mathbf{W} = [\mathbf{w_1} \mathbf{w_2}...\mathbf{w_R}] \in \mathbb{R^{K \times R}}$, then $\mathbf{U}$, $\mathbf{V}$, $\mathbf{W}$ are called the {\em factor matrices}. An equivalent representation of Relation~(\ref{eq:cp_deco}) for the input tensor $\tensor{X}$ w.r.t. its frontal slices $\mathbf{X}_k$ is:

\begin{equation} \label{eq:CP_formula}
    \mathbf{X}_k \approx \mathbf{U} \mathbf{S}_k \mathbf{V}^T
\end{equation}

where $k = 1,...,K$, $\mathbf{U}$, $\mathbf{K}$ are factor matrices and $\mathbf{S}_k \in \mathbb{R}^{R \times R \times K}$ is an auxiliary tensor.

\noindent{\em Interpretability and uniqueness:} CP factors admit to an intuitive interpretation and is provably unique. In Relation~(\ref{eq:cp_deco}), each rank one component represents a latent concept. For each $r$-th concept, the vectors $(\mathbf{u}_r, \mathbf{v}_r, \mathbf{w}_r)$ are considered soft-clustering membership indicators, for the corresponding $I$, $J$ and $K$ elements of each mode. In correspondence with Singular Value Decomposition (SVD) for Relation~(\ref{eq:CP_formula}), CP can be interpreted as follows: each slice $\mathbf{X}_k$ is decomposed to a set of factor matrices $\mathbf{U}$, $\mathbf{V}$ which are common for all the slices. Each frontal slice $\mathbf{S}_k$ of $\mathbf{S}$ is a diagonal matrix whose diagonal elements are the $k$-th row of the third component matrix $\mathbf{W}$.

\noindent {\bf PARAFAC2}

\begin{figure}[t!]
\includegraphics[width = 1\textwidth]{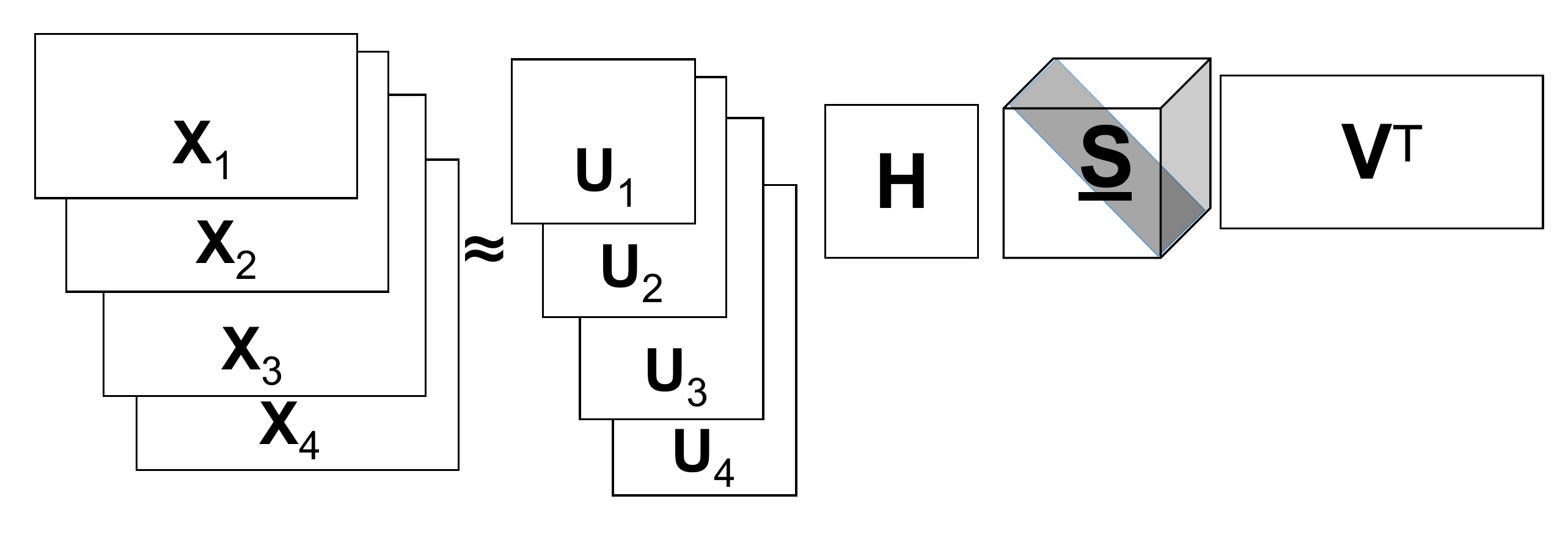}
\caption{PARAFAC2 decomposition of a multi-set}
\label{fig:parafac2}
\end{figure}

\hide{CP decomposition requires strict trilinearity i.e. equal number of rows and columns for each slice in $\tensor{X}$.} Parallel Factorization 2 (PARAFAC2) is a less restrictive model than CP and can successfully deal with multi-set representation of data with an incomparable mode of each slice $\mathbf{X}_k$. It does so introducing a set of $\mathbf{U}_k$ matrices replacing the $\mathbf{U}$ matrix of the CP model in Relation~(\ref{eq:CP_formula}). PARAFAC2 decomposes each slice $\mathbf{X}_k$ as shown in Figure~\ref{fig:parafac2}: 

\begin{equation} \label{eq:parafac2_formula}
    \mathbf{X}_k \approx \mathbf{U}_k \mathbf{S}_k \mathbf{V}^T
\end{equation}

where $k = 1,...,K$, $\mathbf{U}_k \in \mathbb{R}^{I_k \times R}$, $\mathbf{S}_k \in \mathbb{R}^{R \times R}$ is a diagonal and $\mathbf{V} \in \mathbb{R}^{J \times R}$. The cross product of $\mathbf{U}_k^T \mathbf{U}_k$ is invariant regardless of the $k$ involved. This relaxes the CP model's invariance of the factor $\mathbf{U}_k$, thus preserving the uniqueness of the solution. For the above constraint to hold, each $\mathbf{U}_k$ factor us decomposed as: 

\begin{equation}\label{eq:p2_constraint}
    \mathbf{U}_k = \mathbf{Q}_k \mathbf{H}
\end{equation}

where $\mathbf{Q}_K$ is of size $I_k \times R$ and has orthogonal columns, and $\mathbf{H}$ is an $R \times R$ matrix, which does not vary by $k$. Then, the constraint $\mathbf{U}_k^T \mathbf{U}_k$ is constant over $k$ is implicitly enforced, as follows: $\mathbf{U}_k^T \mathbf{U}_k = \mathbf{H}^T \mathbf{Q}_k^T \mathbf{Q}_k \mathbf{H} = \mathbf{H}^T \mathbf{H} = \mathbf{\Phi}$. We use the classical algorithm proposed by Kiers et al. \cite{Kiers1999parafac2} for fitting PARAFAC2, which follows an Alternating Least Squares (ALS) approach and expects dense data as input.

\section{Proposed Method and Experiments}
\label{sec:experiments}

In this section, we present the \method framework and the experiments we conducted using the {\em StudentLife} dataset.  which uses combinations of features at different granularity levels to create privacy surface configurations to predict user behaviors. Given the privacy surfaces, we investigate the utility-intrusiveness spectrum, i.e. the trade-off between the intrusiveness and prediction capability. We compare the effectiveness of privacy surfaces with different levels of intrusiveness. We hypothesize that different privacy surfaces present different states of the user. \hide{In addition, we would also like to know how a set of privacy surfaces with varying degrees of intrusiveness can yield results of acceptable quality in predicting various personality traits of a user.} We explore the commonalities in these user clusters and validate them against the ground truth.

\subsection{\method framework}

\noindent {\bf Step 1: Create privacy surface configurations} \\ 

\begin{figure}[t!]
	\subfigure[Privacy surface configuration with high level of intrusiveness. \label{fig:config_1}]{\includegraphics[width=0.95\textwidth]{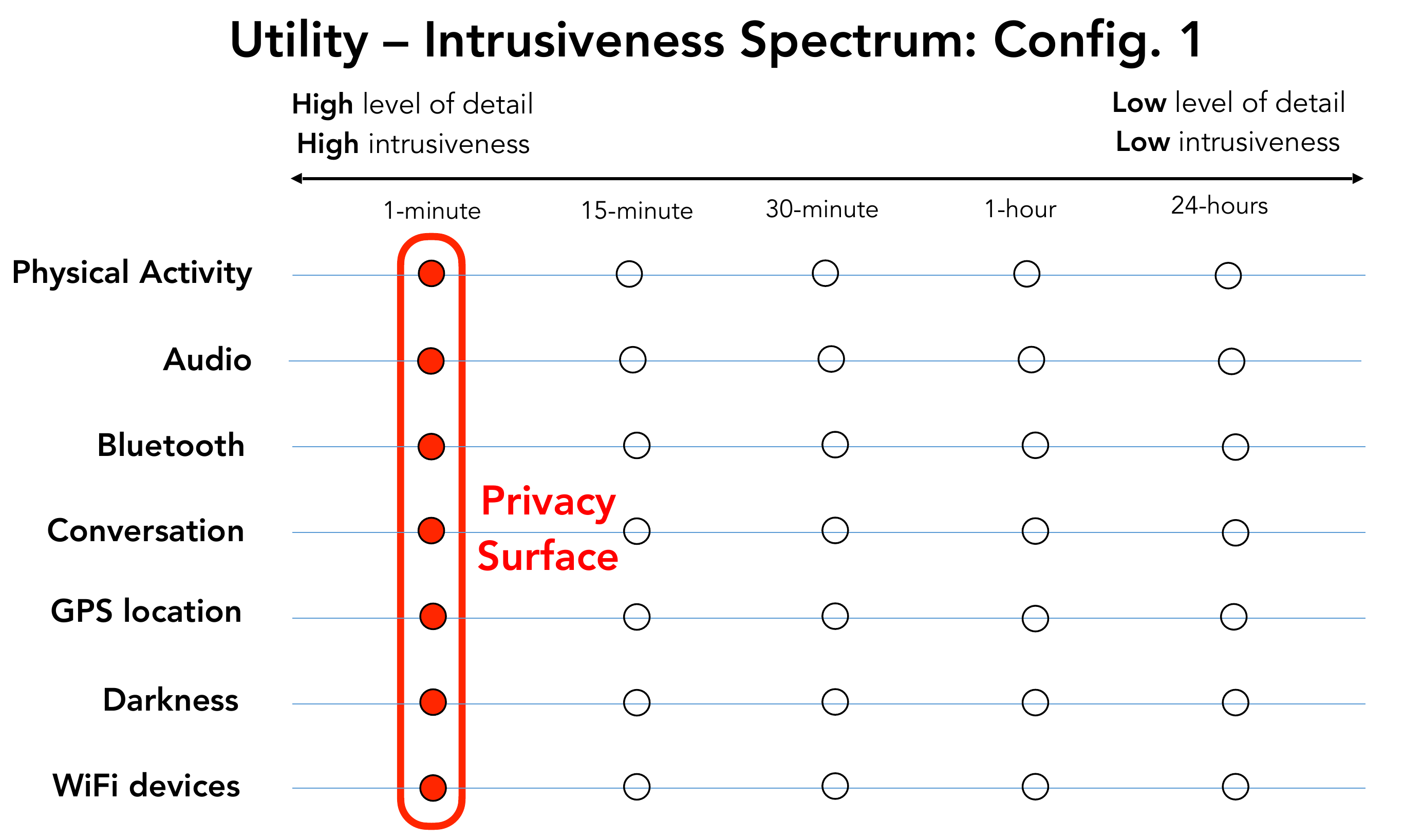}} \\
	\subfigure[Privacy surface configuration with mixture of different levels of intrusiveness, some features with high level of intrusiveness, while some at low level. \label{fig:config_2}]{\includegraphics[width=0.95\textwidth]{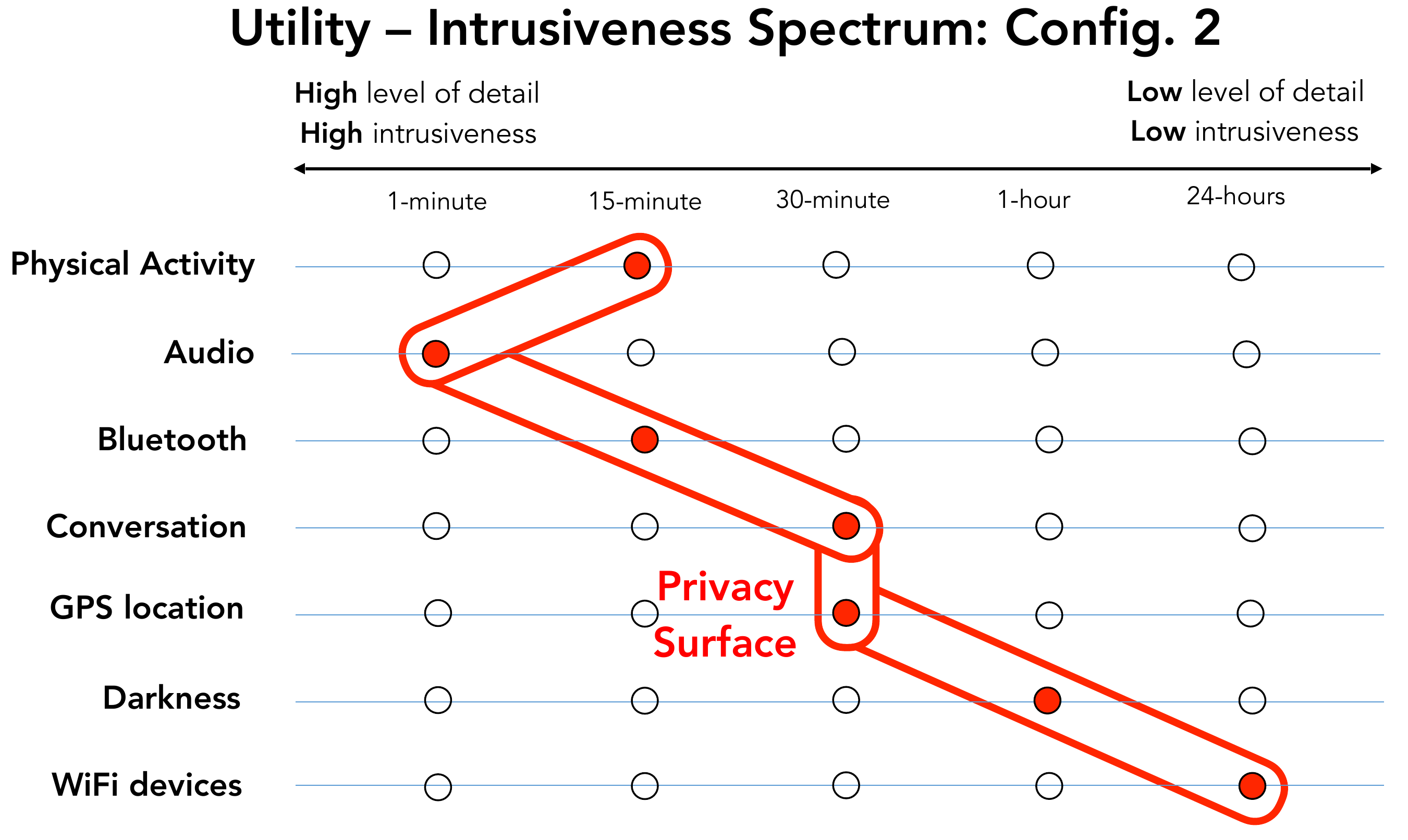}}
	\subfigure[Privacy surface configuration with low level of intrusiveness.\label{fig:config_3}]{\includegraphics[width=0.95\textwidth]{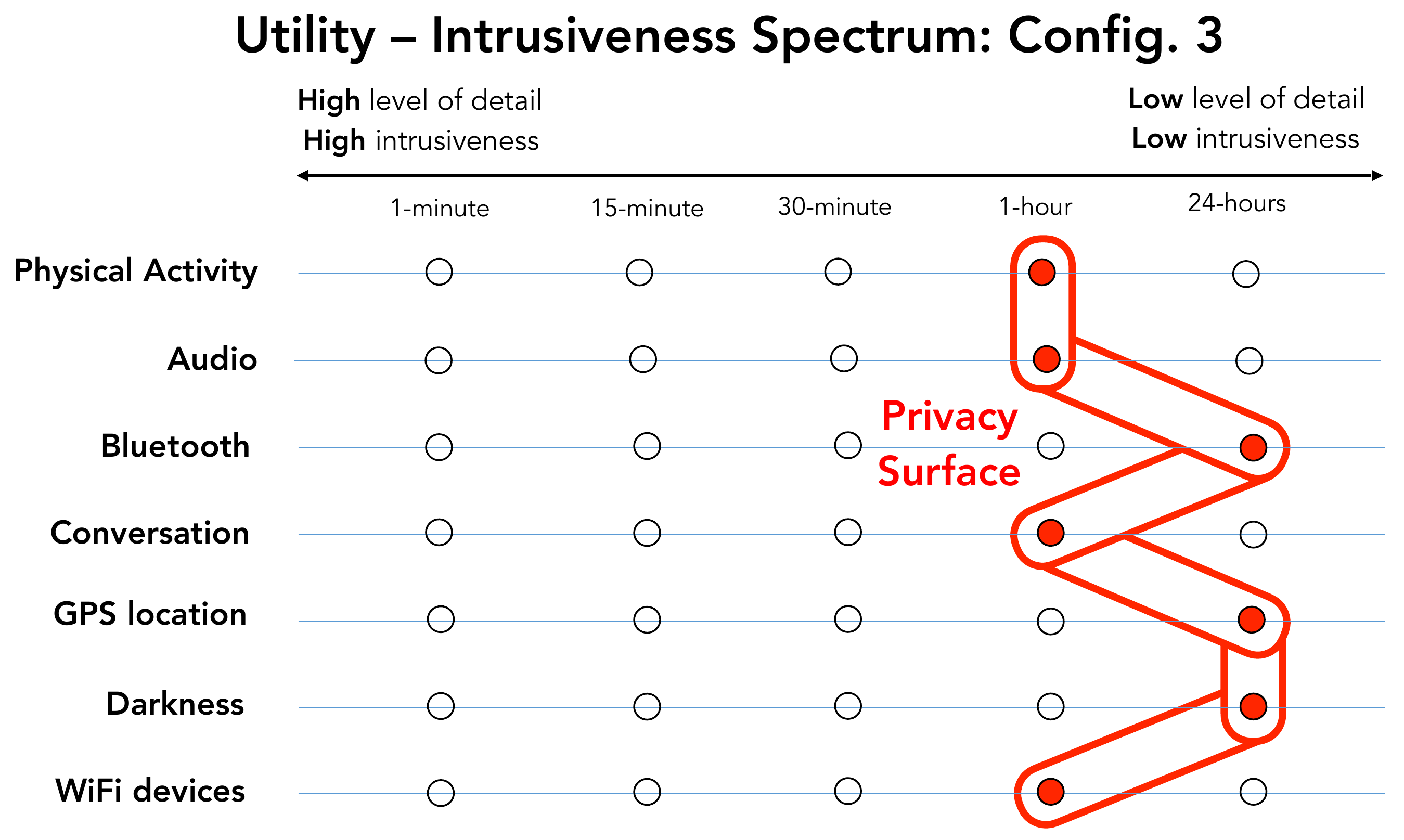}}
	\caption{Three indicative privacy surfaces across the utility-intrusiveness spectrum. The configuration of privacy surface (a) is relatively highly intrusive, and we hypothesize for it to have high utility and effectiveness. Surface (b) has a mixture of in terms of intrusiveness. Surface (c) rates low on the intrusiveness spectrum and we hypothesize that the utility will accordingly be lower. We study and discuss how the utility and effectiveness vary as we move between different privacy surfaces and prescribe different privacy surface configurations based on what to predict, with varying levels of intrusiveness and associated utility.}
\end{figure}

A privacy surface refers to a combination of features each aggregated at varying temporal granularities. \hide{Features with fine temporal granularity have high intrusiveness than those with coarse granularity.} We created three sets of five configurations, a total of 15 different privacy surfaces. As discussed in Section~\ref{sec:problem}, we extracted 18 features from the smartphone sensor data, each aggregated at five temporal granularities, arranged in the order of low to high privacy levels - 1-minute, 15-minute, 30-minute, 1-hour, and 1-day time bins.\hide{, where 1-minute through 1-day represents high to low on the intrusiveness spectrum.}

The first set consists of all the extracted features, each set with the same temporal granularity. We created three-way tensors with modes (time, user, feature), where we have 48 users, 18 features and the length of the time mode is one of the temporal granularities. Figure~\ref{fig:config_1} shows a sample privacy surface with features  aggregated at 1-minute intervals. With expositional clarity from previously published work, we identified features that correlate with mental health states such as depression and stress. For example, the features darkness and silence determine sleep patterns of a user, where insomnia is strongly correlated with depression. Consequently, the second set of privacy surface configurations consist of a reduced set of features that correlate with mental health states. We use these surveys in the validation part of our framework. We created three-way tensors with modes (time, user, feature), with 48 users, 18 features and the length of the time mode is one of the temporal granularities.

The third set of privacy surface configurations consists of combinations of features each at a different temporal granularity, forming a multi-set with modes (time, user, feature). \hide{A multi-set is a collection of slices that is relaxed in one mode. An example of a multi-set structure is presented in the Figure~\ref{fig:tensor_vs_multiset}. In this case, the feature mode is the slice, the columns and the rows of the feature mode represent the user mode and the time mode respectively. Since each feature varies in temporal granularity, the time mode varies per feature.} Example privacy surface configurations are presented in Figures~\ref{fig:config_2} and ~\ref{fig:config_3}. Figure~\ref{fig:config_1} represents the features of all users aggregated into 1-minute bins and we regard it as a highly intrusive privacy surface. Figure~\ref{fig:config_2} has few features with fine granularity while some features have coarse granularity. This represents a mixture of levels of intrusiveness on the utility-intrusiveness spectrum. Figure~\ref{fig:config_3} represents a privacy surface with low level of intrusiveness since most features have coarser temporal granularity.

\noindent {\bf Step 2: PARAFAC2 for unsupervised user clustering} 

We use the PARAFAC2 decomposition presented in Section~\ref{sec:method} to compute user clusters in an unsupervised fashion. The rank $R$ refers to the number of components. The input to PARAFAC2 is a multi-set (time, user, feature) that represents a privacy surface configuration. \hide{Each frontal slice is a feature with columns as users and rows are instances of that feature in time.} We solve the optimization problem in Equation~\ref{eq:parafac2_formula} to decompose the multi-set. Depending on the temporal aggregation,  5\% to 18\% of the multi-set data is missing. PARAFAC2 decomposition handles this by a straightforward iterative imputation of missing data after each full cycle of updates. The goal of the decomposition is to capture clusters with high intra-cluster homogeneity based on their temporal signals. We used PARAFAC2 code from \cite{Perros2017spartan} to compute the user clusters.

\noindent \underline{{\em Rank approximation}}: Before we decompose the multi-set, we need to find the correct number of components, $R$. For a tensor with dimensions $I$, $J$, and $K$, the maximum possible rank is $min(I, J, K)$. We modified the {\sc AutoTen} algorithm proposed by Papalexakis\cite{Papalexakis2016autoten} to find the best approximation for $R$. In a nutshell, {\sc AutoTen} is maximizing the number of latent factors that can be extracted with high quality, and the quality is measured by the so-called ``Core Consistency'', which indicates whether a particular CP decomposition is adequately capturing low-rank structure in the data. The way we modify {\sc AutoTen} is by expressing the PARAFAC2 model in its CP form, since the second step of PARAFAC2 is essentially CP ``slice-wise'' formulation of Relation~\ref{eq:CP_formula} and produces a tensor as an intermediary outcome. We input this tensor to {\sc AutoTen}, which approximates the best $R$ value. The rank lies between 3 and 6 for most privacy surface configurations, in agreement with our general intuition. The {\sc AutoTen} rank for privacy surfaces in Figures ~\ref{fig:config_1}, \ref{fig:config_2}, and \ref{fig:config_3} is 3, 4, and 4 respectively.

\noindent \underline{{\em Model Interpretation}}: The goal behind our clustering is to achieve homogeneity in terms of temporal trends and mental health states per cluster using minimum amount of information. We propose the following model interpretation towards that goal: 
\begin{itemize}
    \item For the common factor matrix $\mathbf{V}$, the non-zero values of each $r$th column indicate the {\em user membership} to the corresponding $r$th cluster. 
    \item The diagonal $\mathbf{S}_k$ is a feature by cluster matrix that provides the {\em importance membership indicators} of the $k$th feature to each one of the $R$ clusters. Sorting the  $R$ columns gives the order of feature importance to each $r$th cluster.
    \item Each $\mathbf{U}_k$ factor matrix provides the {\em temporal signature} of the cluster: each $r$th column of $\mathbf{U}_k$ reflects the temporal evolution of the $r$th cluster with respect the time granularity of feature $I_k$.
\end{itemize}

\hide{
The outcome of PARAFAC2 is users that fall into various clusters based on the rank input. Essentially, the decomposition finds groups similar individuals and finds latent temporal factors in the noisy, heterogeneous data revealing its low-dimensional patterns. 
The constraint imposed in Eq~\ref{eq:parafac2_formula} allows for uniqueness in the interpretation. The $V$ matrix provides the soft clustering membership for the users. We sort the users by their probability of belonging to a cluster and explore the properties of the members in those components.

All experiments were conducted on a 4-core desktop running Ubuntu 14.04 with 64 GB of RAM, each core with a maximum clock frequency 3.70GHz. No hyper-threading was enabled.}

\begin{figure}[t!]
	\subfigure[Set 2: Depression measure variance. \label{fig:exp2_v1}]
	{\includegraphics[width=0.48\textwidth]{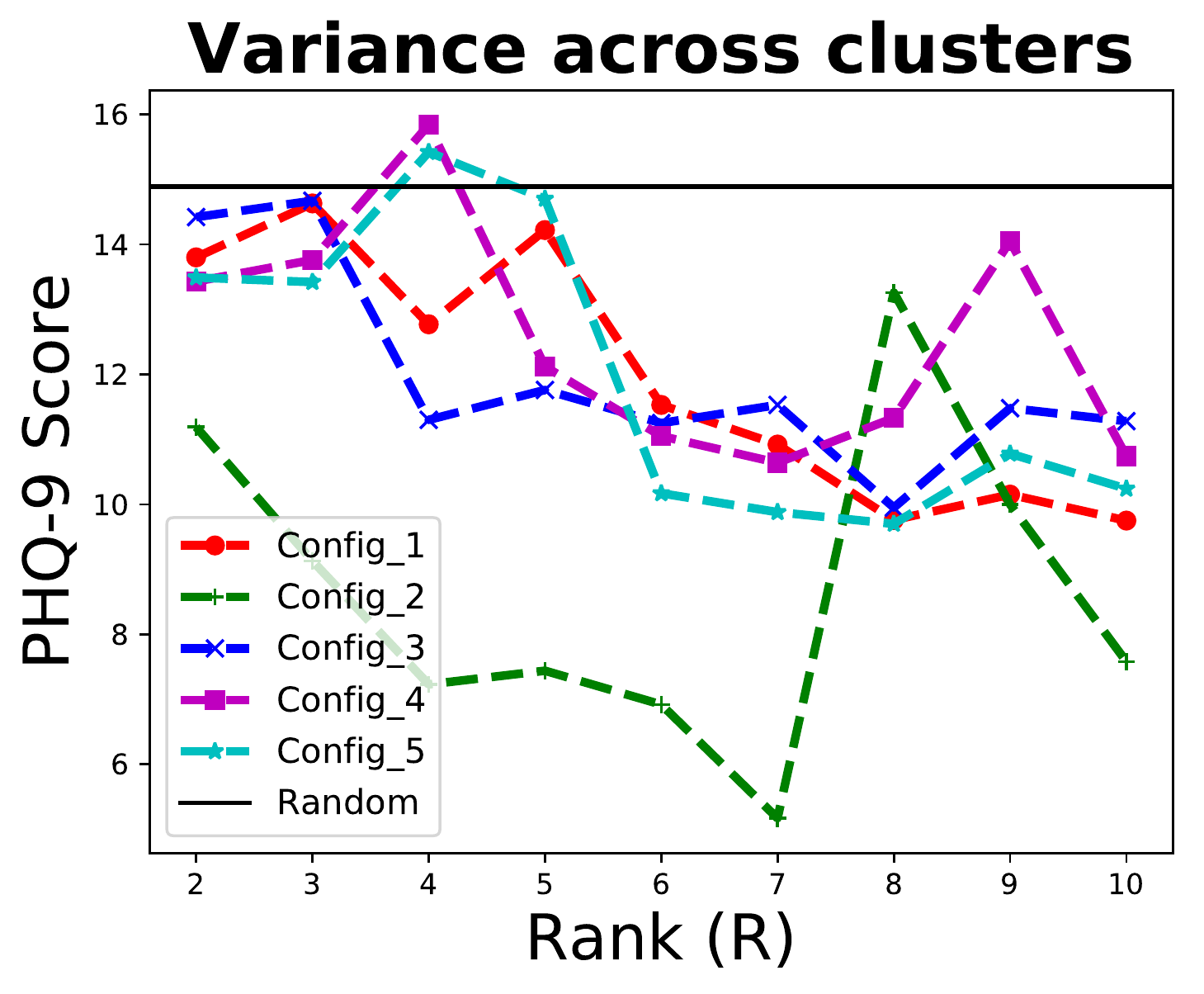}}
	\subfigure[Set 2: Stress measure variance. \label{fig:exp2_v2}]
	{\includegraphics[width=0.48\textwidth]{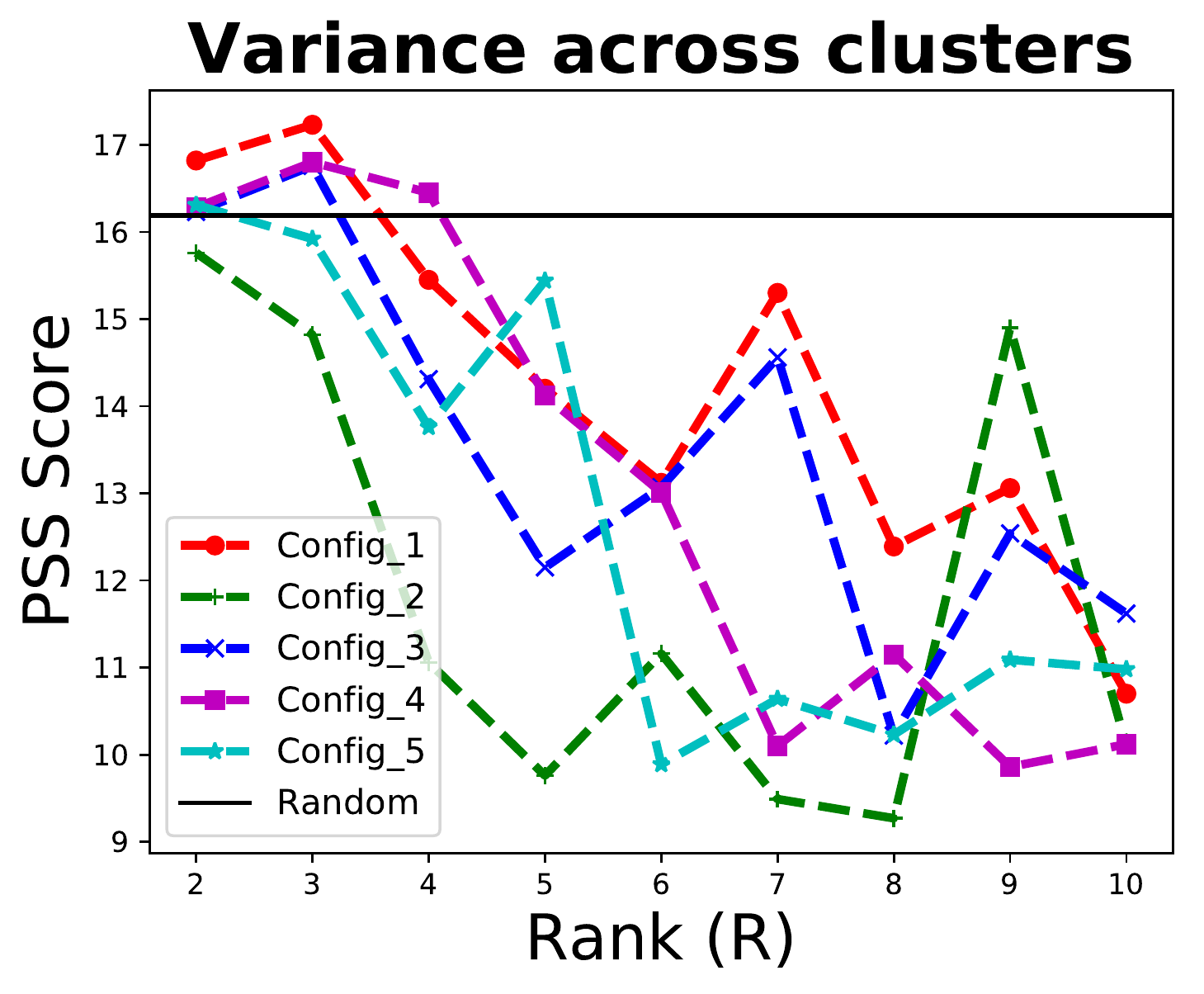}} \\
	
	\subfigure[Set 2: Depression measure IQR. \label{fig:exp2_i1}]
	{\includegraphics[width=0.48\textwidth]{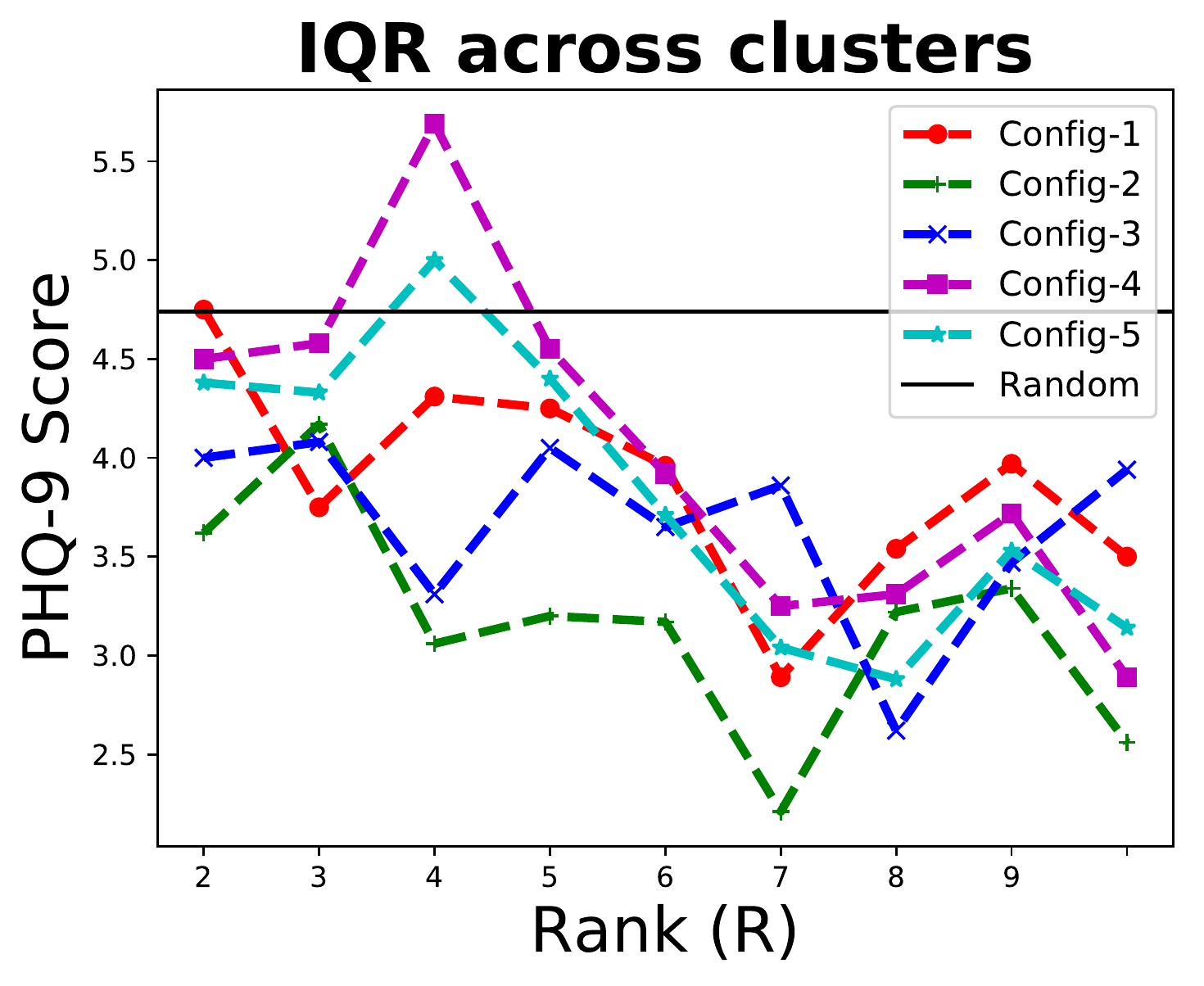}}
	\subfigure[Set 2: Stress measure IQR. \label{fig:exp2_i2}]
	{\includegraphics[width=0.48\textwidth]{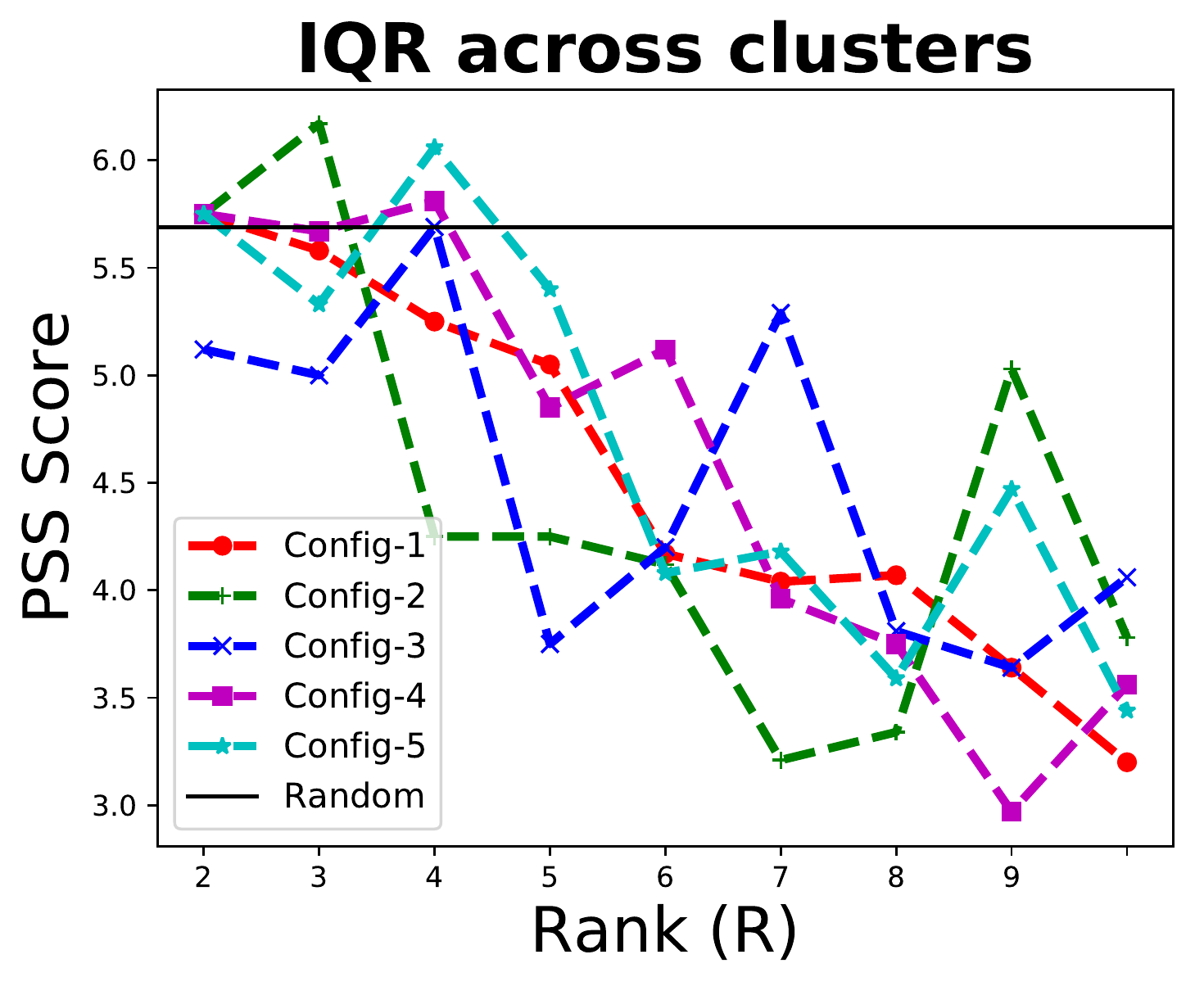}}
	\caption{Mean variance and mean interquartile range of pre-survey measures of depression and stress for clusters in Set 2 configurations. Privacy surface {\em Config-2} has low variance for both measures at ranks 4 and 5 and {\sc AutoTen}'s rank approximation is 4. To ensure non-randomness in outcomes, we compare them against baseline {\em Random} that computes mean variance and mean IQR of 10 randomly picked measures. The baseline varies significantly from the configuration values.}
	\label{fig:exp2_measures}
\end{figure}

\begin{figure}[t!]
	\subfigure[Set 3: Depression score variance.\label{fig:exp3_v1}]
	{\includegraphics[width=0.48\textwidth]{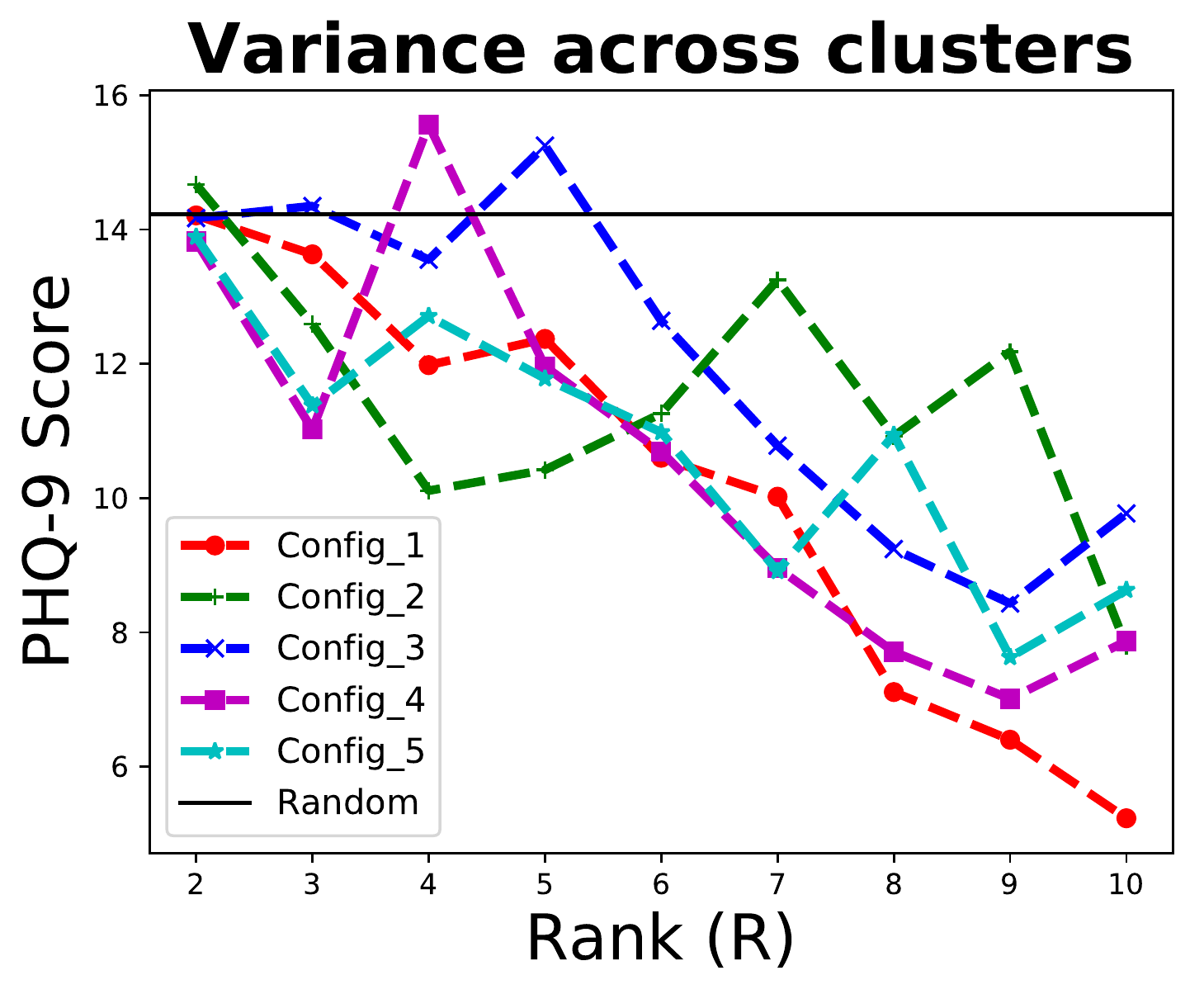}}
	\subfigure[Set 3: Stress score variance.\label{fig:exp3_v2}]
	{\includegraphics[width=0.48\textwidth]{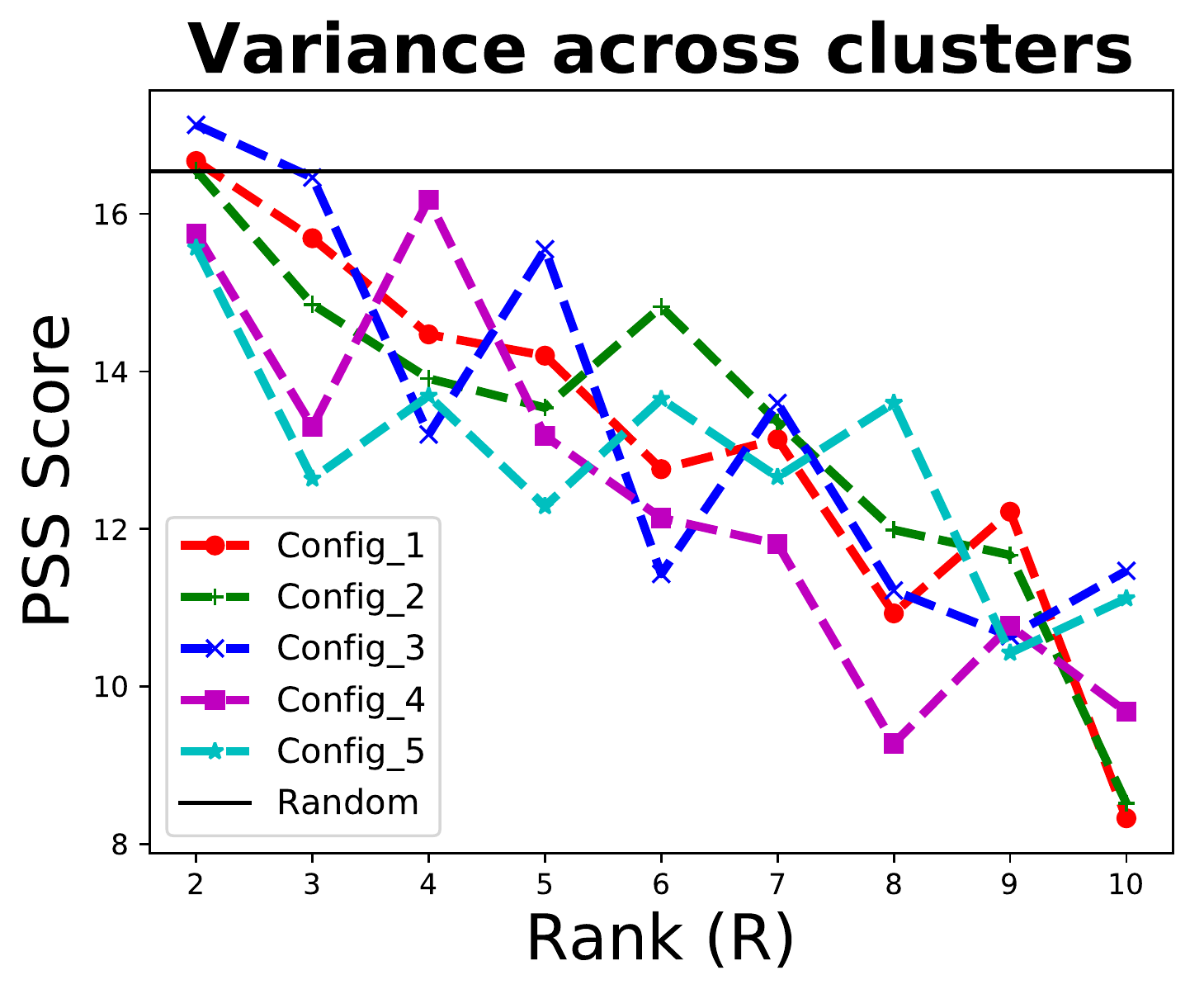}}
	\\
	\subfigure[Set 3: Depression score IQR. \label{fig:exp3_i1}]
	{\includegraphics[width=0.48\textwidth]{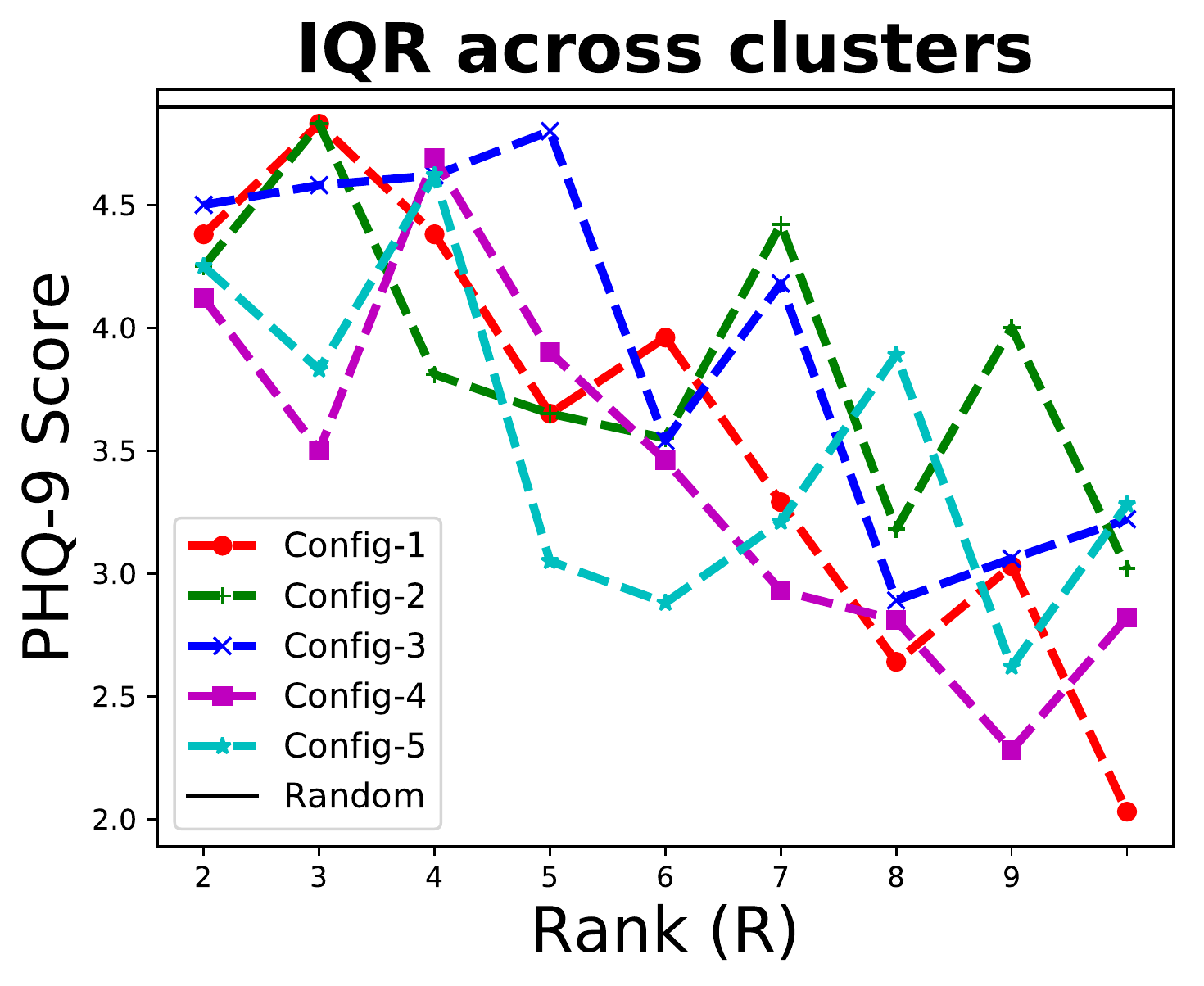}}
	\subfigure[Set 3: Stress score IQR. \label{fig:exp3_i2}]
	{\includegraphics[width=0.48\textwidth]{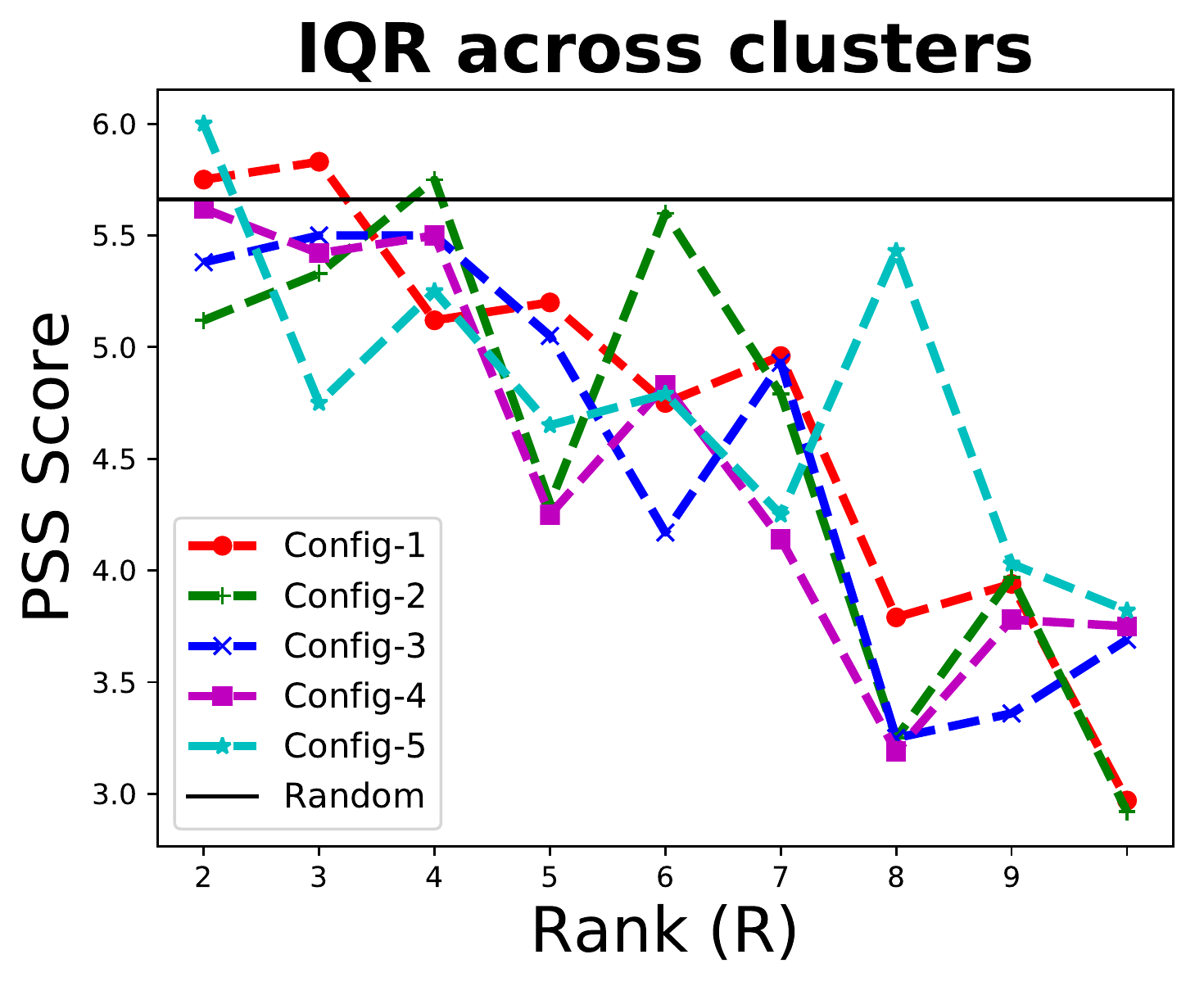}}
	\caption{Mean variance and mean interquartile of pre-survey measures of depression and stress for clusters in Set 3 configurations. Privacy surfaces {\em Config-1} and {\em Config-2} have low variance for both measures at ranks, 3, 4 and  5, for which {\sc AutoTen}'s best approximation is 5 and 4 respectively. The baselines varies significantly from the Config values.}
	\label{fig:exp3_measures}
\end{figure}

\hide{
\begin{table}[t]
    \centering
    \begin{tabular}{c|c|c|c|c|c}
          & State 1 & State 2 & State 3 & State 4 & State 5\\
         \hline \hline
         Config 1 & \cmark & \xmark & \cmark & \xmark & \cmark \\
         Config 2 & \xmark & \cmark & \cmark & \cmark & \xmark \\
         Config 3 & \cmark & \xmark & \xmark & \xmark & \cmark \\
         \hline
    \end{tabular}
    \caption{Accuracy of mental health state clustering for sample privacy surface configurations}
    \label{tab:config_state}
\end{table}
}

\subsection{Validation using psychometric scales} 

We analyze the components of the soft membership matrix $\mathbf{V}$ obtained from PARAFAC2 decomposition and consider the top-2 cluster memberships for each user. We hypothesize that members in a cluster share a similar mental state and temporal evolution. From the list of mental health states in Table~\ref{tab:surveys}, we chose pre-survey measures of depression and stress to present in our plots. Lower scores indicate low levels of depression and stress. As a validation for rank computed using {\sc AutoTen}, we compute mean variance and mean interquartile range of mental health states for users in the cluster for all privacy surface configurations. With this, we can observe configurations with low variance and their respective statistical dispersion. Figures~\ref{fig:exp2_measures} and \ref{fig:exp3_measures} present mean variance and mean IQR for privacy surfaces in set 2 and set 3. To ensure that the clustering is non-random, we compare these data points against a baseline that picks 10 random values from the survey measures and computes their mean variance and mean IQR. 

The mean variance of depression score for configurations in Set 2 presented in Fig~\ref{fig:exp2_v1} is lower for {\em Config-2} at rank 4, while the mean IQR in Fig~\ref{fig:exp2_i1} for depression lies in the range 3 and 4. This is the same observation in stress measure as well presented in Figs~\ref{fig:exp2_v2} and \ref{fig:exp2_i2}. Figure~\ref{fig:exp3_measures} presents the mean variance and mean IQR for configurations in Set 3. The configurations {\em Config-1} and {\em Config-2} in Fig~\ref{fig:exp3_measures} are the same privacy surfaces represented in figures~\ref{fig:config_2} and \ref{fig:config_3}. Fig~\ref{fig:exp2_v1} exhibits low mean variance for {\em Config-2} at rank 4 and the mean IQR lies between 3 and 4. This observation are the same as for the privacy surface with high intrusiveness shown in Fig~\ref{fig:exp2_measures}. These plots empirically validate that we can create privacy surfaces with low level of intrusiveness and high utility on the utility-intrusiveness spectrum and form homogeneous clusters high privacy level temporal data.

\subsection{Exploratory analysis of discovered clusters}

\begin{figure}[t!]
	\subfigure[Per-cluster depression score.\label{fig:exp3_b1}]
	{\includegraphics[width=0.48\textwidth]{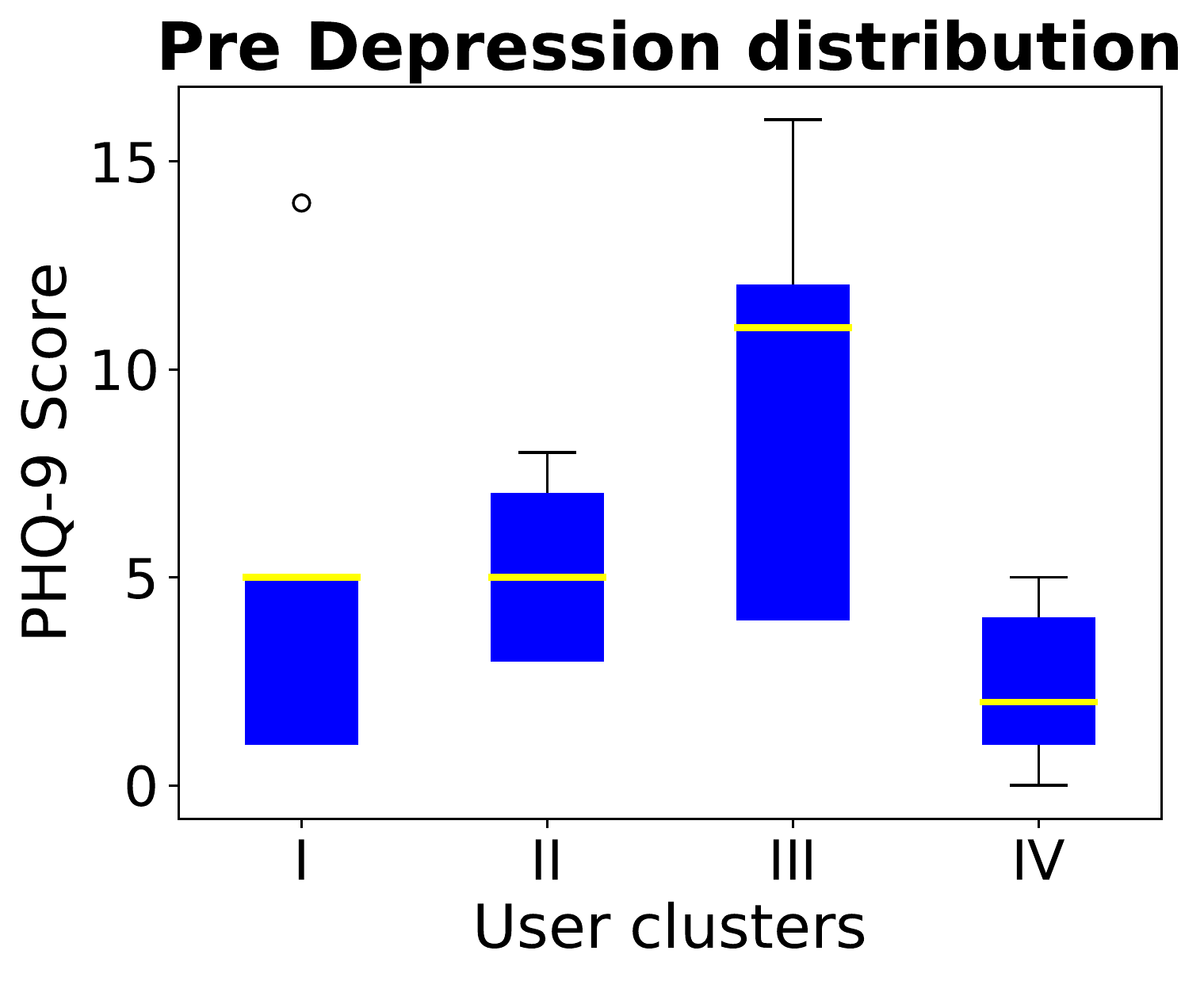}}
	\subfigure[Per-cluster stress score .\label{fig:exp3_b2}]
	{\includegraphics[width=0.49\textwidth]{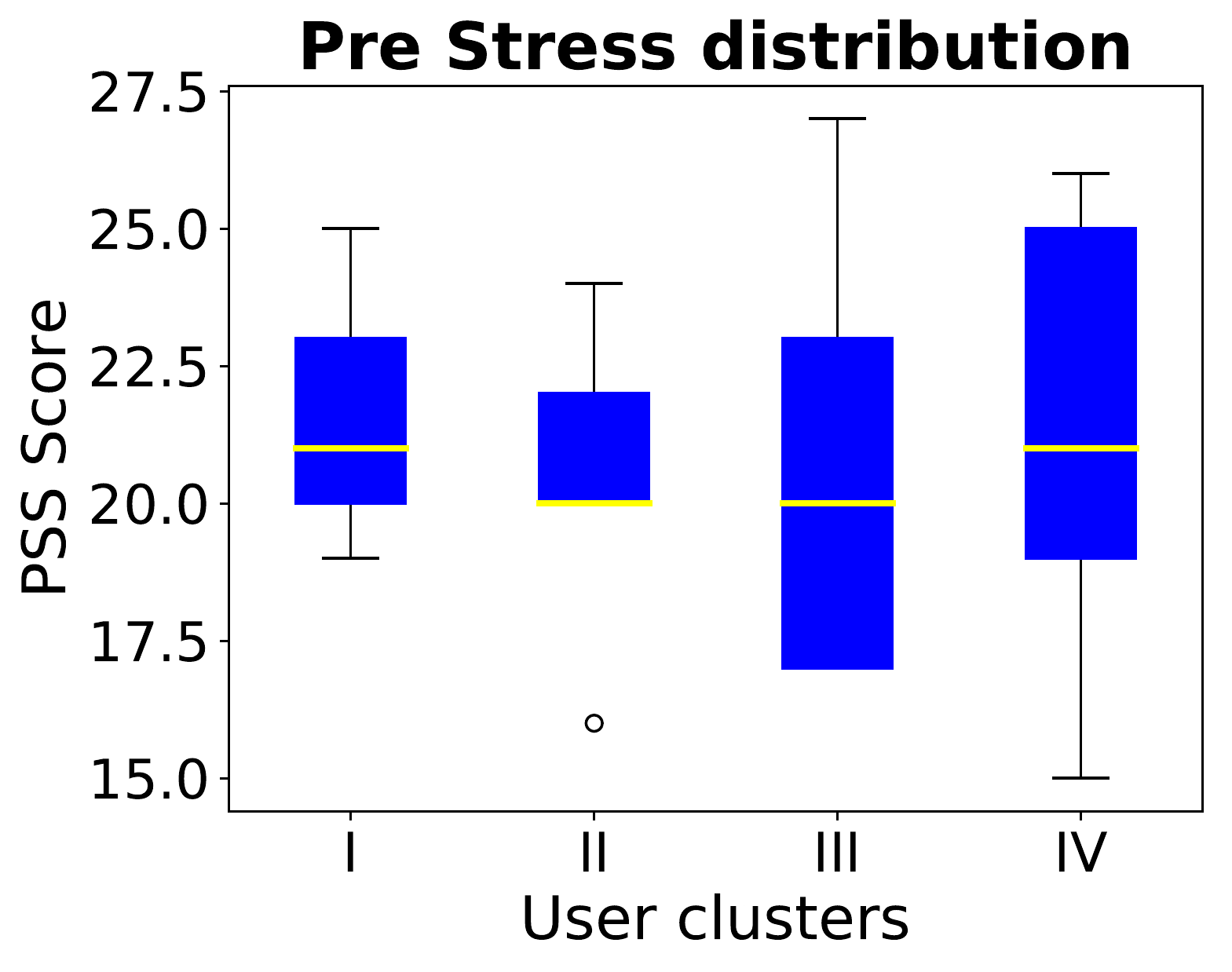}} \\
	\caption{Box plot pre-survey depression and stress measures of top-5 users in Component 1 for privacy surface {\em config-3}, computed using PARAFAC2.}
	\label{fig:box_plots}
\end{figure}

\begin{figure}
    \centering
    \includegraphics[width=1\textwidth]{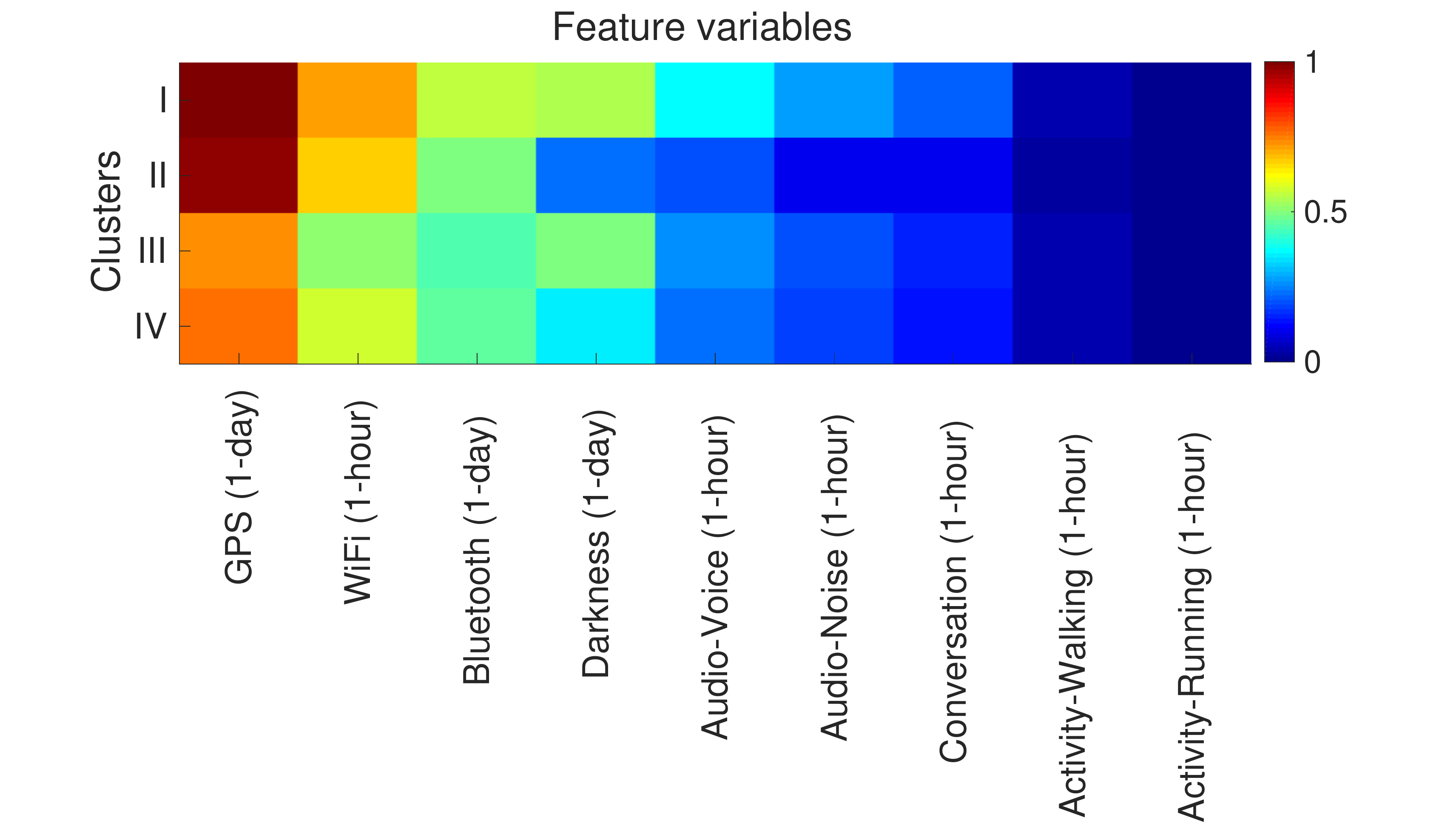}
    \caption{Membership of variables in each of the four components for Config~3 from the factor matrix $\mathbf{S}_k$.}
    \label{fig:feature_membership}
\end{figure}

We present an in-depth analysis on user clusters discovered using PARAFAC2 decomposition on privacy surface configuration in  Figure~\ref{fig:config_3}, where all features have a temporal granularity of either 1-hour or 1-day. The factor matrix $\mathbf{V}$ indicates the likelihood of user membership in each $r$th cluster. Figure~\ref{fig:box_plots} presents plots to observe the dispersion of sample mental health states of pre-survey depression and stress for top-10 users in each cluster. The inter-cluster variation in Figure~\ref{fig:exp3_b1} is high with different median values visibly different for each cluster and the intra-cluster variation is low showing that the clusters significantly vary from one another. The mean depression scores for clusters 1 through 4 in Figure~\ref{fig:exp3_b1} is 6.14, 5.6, 7.32 and 2.17 respectively, while the mean stress scores are 23, 20.6, 21 and 25. Figure~\ref{fig:feature_membership} shows the  importance membership of sensor features in the clusters. The top-3 important features are GPS, WiFi, and Bluetooth for which we consider the count of unique values of (latitude, longitude), \textit{bssid} and device id respectively. These features characterize the physical mobility of users. To understand the temporal evolution of behavior in the clusters, we use the academic information of the users - deadlines, piazza class forum participation, and overall GPA.

\begin{figure}[t]
	\includegraphics[width=1\textwidth]{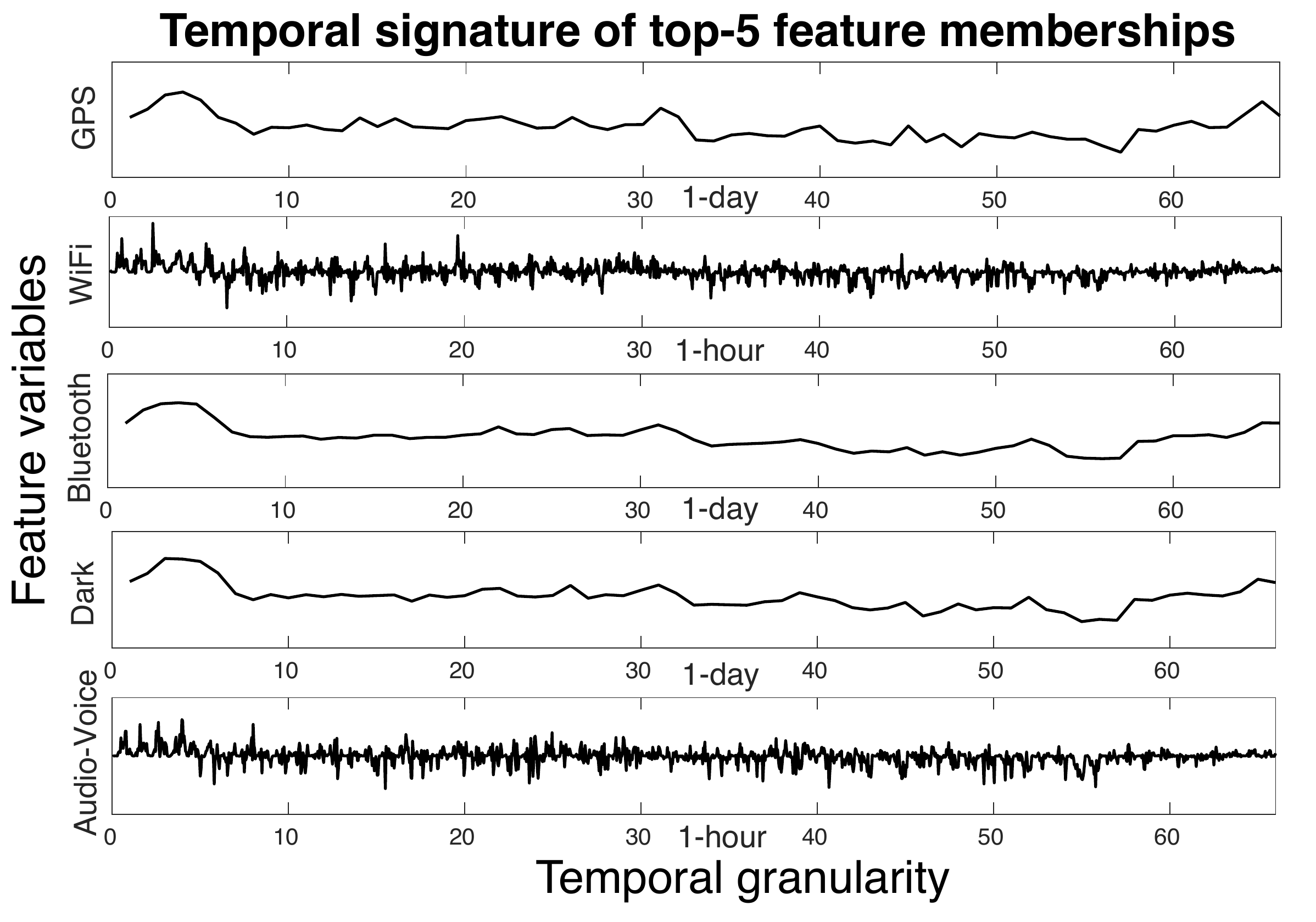}
	\vspace{-0.5cm}
	\caption{Temporal evolution of Cluster 1 features in Config~3 computed using PARAFAC2. The top-5 features in the cluster contain different granularities - GPS, WiFi, Bluetooth, Darkness,  and Audio-voice, with WiFi and Voice at a temporal granularity of 1-hour and the rest at 1-day.}
	\label{fig:c1_time_series}
\end{figure}

Cluster 4 has an highest mean GPA of 3.56, followed by clusters 3, 2 and 4 with GPA 3.45, 3.42 and 3.18 respectively. This places cluster 4 as the top academic performers, who also exhibit the lowest mean depression score among the cluster. Cluster 3 has mildly high self-reported depression and stress, but still maintain a good overall GPA. Figure~\ref{fig:c1_time_series} presents the temporal evolution of the top-5 features in Cluster 1. We correlated the temporal evolution of features in each cluster with the deadlines of users and observed the following: cluster 1 displays a negative correlation (r=-0.3, p=0.01). That is, when there are several deadlines, the members is have less mobility and less sleep. This can be seen from a raise in the peak at the beginning of the quarter and the fall in Figure~\ref{fig:c1_time_series} at day 34, where members had the highest number of deadlines. This is also the case with cluster 4, while the temporal evolution of cluster 2 and 3 exhibit only weak negative correlation with deadlines. Clusters 1 and 3 have the highest rate of participation in the Piazza online forum while cluster 2 users has the lowest participation rate. Cluster 1 and 4 have the highest mean flourishing score pre-survey, which measures the perceived success and it raises higher in the post-survey values. However, cluster 1 has the highest negative affect pre-survey and it reduces post-survey. Cluster 3 displays the highest loneliness both pre- and post-survey and coincidentally, they have also have lower academic success as seen from GPA and class participation. This makes us think there is a strong correlation between a member's self-reported loneliness and them thriving academically. We believe that clustering the privacy surface Config 3 using PARAFAC2 has presented us with high-quality clusters where each cluster displays a set of properties with low levels of intrusiveness. We validate and explore the traits of cluster members using the ground truth provided in the StudentLife dataset.

\section{Related Work}
\label{sec:related}

Our work lies in the intersection of various topics that researchers have extensively studied in the past. We discuss the state of the art of each of those areas in the next few lines. However, we must point out that to the best of our knowledge, our viewpoint in conducting such an analysis on a multi-dimensional, multi-granular smartphone sensing data is to create a privacy-aware framework that uses minimal amount of information to derive behavioral patterns. 

\subsection{Online Behaviors and Mental Health}

Online social networks have been used to examine \hide{friendship groups \cite{Gjoka10friendship},} interactions between users \cite{Devineni15powerwall} and predicting attributes like age, gender, ethnicity, and political affiliations \cite{Kosinski13traits}. In the recent years, user activity on social media has widely been used to understand mental health behaviors like PTSD, depression, bipolar disorder \cite{coppersmith14signals} and suicidal ideation \cite{Dechoudhury16suicidal}. In our previous work, we studied changes in temporal behaviors of users and their relationship to real-life events \cite{Devineni17inet}. Smartphones come equipped with the sensors that let researchers continually collect data. Understanding temporal patterns of stability and change can provide insight into individual differences, mental health, and real-life events \cite{Wang2014studentlife}.

\subsection{Tensors and Multi-set Decomposition}
Papalexakis et al. \cite{Papalexakis2017tensors} presented the most recent survey of tensors, algorithms and its applications. The most efficient algorithm for PARAFAC2 appeared in \cite{Kiers1999parafac2}. \cite{hosseinmardi2018discovering} used CP decomposition to discover low-dimensional structures in smartphone data and found studying, partying and academic deadline patterns in college students. Multi-set analysis and PARAFAC2 have been used very successfully in Chemometrics \cite{Bro1999parafac2} because it allows the joint analysis of chemical measurements that may contain time shifts. Chew et al. \cite{Chew2007lang} used PARAFAC2 to analyze a set of term-document matrices of the same documents for different languages; Because each language has a different set of terms used, it is beneficial to jointly analyze the data into a common reduced dimension. Most recent work uses multi-set mining \cite{Perros2017spartan} to for mining electronic health records of patients and deriving time-evolving phenotypes. 

\subsection{Privacy-aware Mining Frameworks}
The notion of privacy used in our work is related to the aggregation of different sources of information from the user data. There is a wealth of privacy preserving related data mining literature \cite{Agrawal2002hippocratic, Machanavajjhala2007diversity}. \cite{Agrawal2001privacy} presents a detailed survey of data mining algorithms including an expectation maximization algorithm for quantifying perturbed data used for electronic health records. 

\section{Conclusions}
\label{sec:conclusions}

Rich multi-modal user activity data allows researchers to understand low-dimensional patterns while preserving user privacy. In our paper, we proposed the concept of privacy surfaces that combine modalities with different granularities to preserve privacy based on the utility-intrusiveness spectrum. Our proposed method \method  takes these privacy surfaces as input and employs multi-set decomposition in order to compute user clusters, which  are homogeneous with respect to the self-reported mental health measures of the students. This demonstrates that we can preserve privacy while achieving high-quality clustering, as a step towards privacy-preserving analytics in an era of sensitive user information.

\section{Acknowledgements}

The authors would like to thank Homa Hosseinmardi at USC for discussions on tensors and creating the dataset.

\small{
\bibliographystyle{IEEEtran}
\bibliography{prav_refs}
}

\balance

\end{document}